\documentclass[11pt]{article}
\usepackage{geometry}  
\usepackage[parfill]{parskip}     
\usepackage{graphicx}
\usepackage{amsmath}
\usepackage{tikz}
\usepackage{algorithmic}
\usepackage{algorithm}
\usepackage{subfigure}

\def\degree{{k}}
\def\linksRC{k^+}

\def\richClubRank{{\Phi}}

\def\degree{k}

\def\stepFunct{w}
\def\eProb{p}
\def\averageDegDeg{k_{\rm n n}}
\def\entropy{S}
\def\cofVar{c}
\def\inversePartRat{I}

\begin{document}

\begin{center}
{\LARGE Ensembles based on the Rich--Club and how to use them to build soft--communities.\\[1ex]}
{R. J. Mondrag\'on\\[2ex]
Queen Mary University of London\\
School of Electronic Engineering and Computer Science\\
Mile End Road\\
E1 4NS, London\\
email: {\tt r.j.mondragon@qmul.ac.uk}\\
}
\end{center}
\section*{Abstract} 
Ensembles of networks are used as null--models to discriminate network structures.  We present an efficient algorithm, based on the maximal entropy method to generate network ensembles defined by the degree sequence and the rich--club coefficient. The method is applicable for unweighted, undirected networks. The ensembles are used to generate correlated and uncorrelated null--models of a real networks. These ensembles can be used to define the partition of a network into soft communities.

\section{Introduction}
Attention has recently been directed to the generation of network ensembles as they are used to identify statistically significant patterns on real networks~\cite{Bianconi08a,annibale2009tailored,squartini2011}.  The ensemble is used as a reference null--model for comparison purposes. The definition of \emph{communities} is an example of a pattern defined using a null--model. The communities are defined as the groups of nodes where the number of links between the communities is smaller than expected~\cite{newman2004finding}. This `expected number of links between the communities' is evaluated using a null--model.  A well defined null--model should capture the essence of the network's topology without introducing spurious or unwanted structural properties. The most common criterion is to define a null--model that has the same degree sequence as the real network and any other topological property has a nonspecific structure.  This criterion captures the basic topological composition of a network but in some cases it cannot remove higher topological properties due to structural constraints~\cite{boguna2004,Maslov2004}. 

Shannon's entropy can be used to describe our state of knowledge of the network structure~\cite{sole2004information,Bianconi08a,dehmer2011history}. The maximisation of this entropy can be used to create an ensemble that is `maximally non--committal', that is given some constraints, the ensemble is as unbiased as it is possible. These unbiased  network ensembles can be evaluated via the Maximum Entropy method (MaxEnt)~\cite{Bianconi08a,Bianconi08b,Bianconi09,annibale2009tailored,Mondragon2014a} or the Maximum Likelihood method~\cite{squartini2011,squartini2011randomizing,squartini2011randomizingB,squartini2014unbiased}, which are equivalent. The generation of null--models using these information--based methods is attractive,  as they provide analytical expressions to describe the basic properties of the ensemble, and in general, can be computed efficiently. 

In this paper the main aim is to present an algorithm to create null--models defined by the conservation of the degree sequence and the rich--club connectivity. The networks considered here are undirected, unweighted and without self loops but multiple links between a pair of nodes are allowed. The algorithm is based on a previous result where the maximal entropy approach (MaxEnt) was used to generate the network ensemble.  The method is attractive because the set of probabilities describing the ensemble can be evaluated very efficiently~\cite{Mondragon2014a}. We use the method to construct ensembles with similar correlations as the real network and also to construct ensembles with minimal correlations. As a sub-aim we present an example of how to combine these ensembles to define `soft' network--communities.

\section{Generation of the ensembles}
The null--model is obtained using the maximal entropy method with the constraints that the degree sequence and rich--club coefficient are conserved~\cite{Mondragon2014a}. The constraints are the sequences $\{\degree_1 ,\ldots,\degree_N\}$ and $\{\linksRC_1,\ldots, \linksRC_N\}$, where $N$ is the number of nodes. The first sequence contains the degree $\degree$ of the nodes. The nodes are ranked in decreasing degree order, the node with the highest degree is ranked first and so on.  The second sequence is the number of links $\linksRC_r$ that node $r$ shares with nodes $r'$ of higher rank, $r' < r$. The total number of links in the network is $L = 1/2\sum_{r=1}^N \degree_r=\sum_{r=1}^N\linksRC_r$. The ranked based rich--club coefficient~\cite{zhou04} is $\richClubRank_r = (2/(r(r-1))) \sum_{i=1}^r \linksRC_i$, thus conserving $\linksRC_i$ is equivalent to the conservation of the rich--club coefficient. 

\begin{figure}
\begin{center}
\subfigure[]{
\includegraphics[width=6cm]{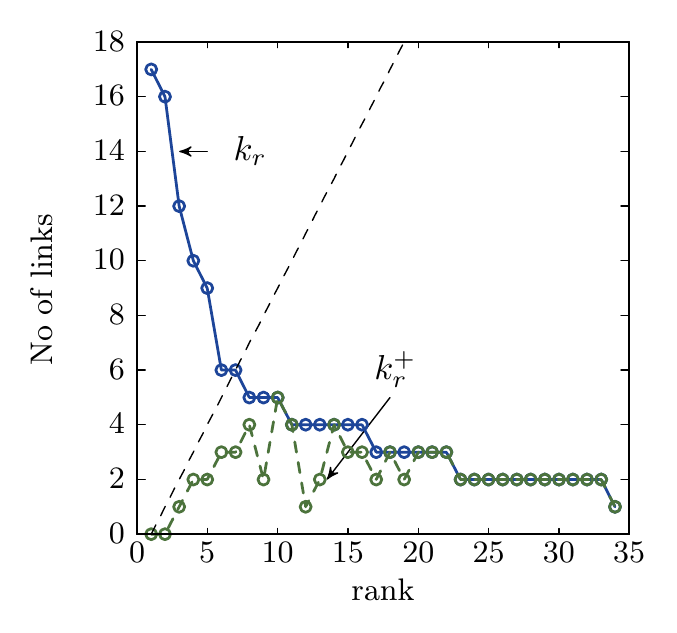}
}
\subfigure[]{
\includegraphics[width=6cm]{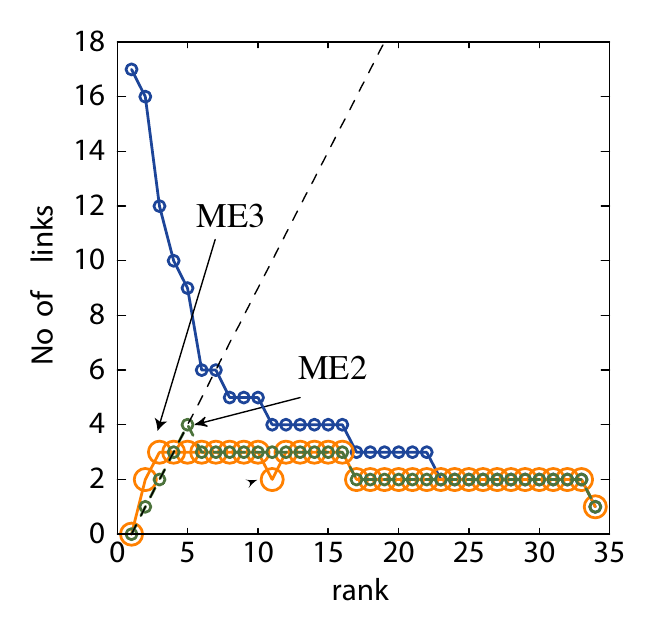}
}
\end{center}
\caption{\label{fig:one} Zachary karate club. (a) The degree $\{\degree_r\}$ sequence (blue) and $\{\linksRC_r\}$ sequence (green), $\linksRC_r$ is always less or equal to $\degree_r$. The dotted line is the bound to restrict the ensemble connectivity to only one link per pair of nodes. (b) The  $\{\linksRC_r\}$ sequence for the null--model with the constraint that on average there is only one link between two nodes (green) and for the null model without this restriction (orange).} 
\end{figure}

The sequence $\{\linksRC_r\}$ is bounded $\linksRC_r\le \degree_r$, but $\linksRC_r$ could be large enough to allow multiple links between two nodes.  If only one link can exist between two nodes then $\linksRC_r\le r-1$. If $\linksRC_r> r-1$ implies that there are more than $r-1$ links between node $r$ and the $r-1$ nodes of higher rank, which means the existence of at least one multiple link (Fig.~\ref{fig:one}(a)).

The Shannon entropy of the network is $\entropy = -\sum_{i=1}^N \sum_{j=1; j\ne i}^N \eProb_{i,j} \log( \eProb_{i,j} )$. The maximal entropy solution is given by the probabilities~\cite{Mondragon2014a}
\begin{equation}
\label{eq:mainResult}
\eProb_{i,j} = 
\frac{\stepFunct(i)\left(\degree_i-\linksRC_i \right) }{\sum_{n=1}^{j-1} \stepFunct(n)\left(\degree_n-\linksRC_n \right)}\frac{ \linksRC_j}{L}\quad i<j\\
\end{equation}
where 
\begin{equation}
\label{eq:stepFunct}
\stepFunct(m) = \frac{\stepFunct(m-1)\sum_{i=1}^{m-1}\stepFunct(i)(\degree_i-\linksRC_i)}{\sum_{i=1}^{m-1}\stepFunct(i)(\degree_i-\linksRC_i)-\linksRC_m\stepFunct(m-1)}.
\end{equation}
The values of $\stepFunct(m)$  are defined recursively with the initial condition $\stepFunct(1)=1$. The average number of links between nodes $i$ and $j$ is $e_{ij}=Lp_{ij}$ with variance $s_{ij}=Lp_{ij}(1-p_{ij})$.

By construction the ensemble satisfies the `soft' constraints $\langle \degree_r\rangle = \sum_{j=1}^N L \eProb_{r, j}=\degree_r $ and $\langle \linksRC_r\rangle = \sum_{j=1}^{r-1} L\eProb_{r, j}=\linksRC_r $, where the variance of the degree is $\sigma^2(\degree_r)=L\sum_{j\ne r}^Np_{r,j}(1-p_{r,j})$. 

We consider three different null--models all constructed using Eqs.~(\ref{eq:mainResult})-(\ref{eq:stepFunct}). Given the degree sequence $\{\degree_i\}$ the nulls are define by the following constraints: 
\begin{enumerate}
\item The sequence $\{ \linksRC_i\}$ is given. We refer to this as the maximal--entropy model ME1.
\item The sequence $\{\linksRC_i\}$  is not given. This sequence is obtained numerically by maximising the entropy with the restrictions that $\linksRC_r\le\degree_r$ and $\linksRC_r\le r-1$. The last restriction is to impose that on average there is only one link between two nodes. We called this model ME2.
\item The sequence $\{\linksRC_i\}$  is not given. This sequence is obtained numerically by maximising the entropy with the restriction that $\linksRC_i\le\degree_i$. In this case it is possible to have, on average, more than one link between a pair of nodes but self--loops are not allowed. We called this model ME3.
\end{enumerate}
The algorithm to evaluate the second and third cases is in the Appendix. Fig.~\ref{fig:one}(b) shows an example of the $\{\linksRC_r\}$ sequence found by our algorithm for the ME2 and ME3 models for the Zachary club.

It is known that the rich--club coefficient is related to the degree--degree correlation~\cite{Colizza06}. As one of the ME1 ensemble constraint is the conservation of the rich--club sequence, the degree--degree correlation produced by the ensemble is similar to the degree--degree correlation of the real network~\cite{Mondragon2014a,mondragon2012random}.

The ME2 removes the restriction that the ensemble should have the same rich--club sequence as the real network but still has the restriction that on average there should be only one link per pair of nodes. The consequence of this last restriction is that 
the ME2 model will produce correlated ensembles if the maximal degree $\degree_{max}$ is greater than the cut--off degree $\degree_c=\sqrt{2L}$~\cite{Maslov2004,boguna2004}.

 The ME3 removes the restriction that there is only one link per pair of nodes but self loops are not allowed. This restriction should produce decorrelated or almost decorrelated networks~\cite{boguna2004}. 

To decide if the ensemble generates a biased null--model we use the average nearest  neighbours degree  given by $\langle k_{\text{nn}}(k) \rangle =\sum_{k'} k' P(k'|k)$~\cite{Pastor01}, where $P(k'|k)$ is the conditional probability that given a node with degree $k$ its neighbour has degree $k'$. For an uncorrelated network $\langle k_{\rm nn} \rangle = {\langle k^2 \rangle}/{\langle k \rangle}$. In our case $\eProb_{i,j}$ is known  so
\begin{equation}
\langle k_{\text{nn}}(k)\rangle = \frac{1}{N_k}\sum_{i=1}^{N}\left(\frac{1}{k}
\sum_{j=1}^{N} \eProb_{i,j}L k_j\right)\delta_{k_i,k},
\end{equation}
where $\delta_{k_i,k}=1$ if $k_i=k$ and zero otherwise.

Another approach to evaluate if the statistical properties of the ensemble are biased is via the \emph{coefficient of variation} and the \emph{inverse participation ratio} for the degree sequence~\cite{squartini2014unbiased}.
The coefficient of variation for the degree $\degree_i$ is 
\begin{equation}
\cofVar(\degree_i) = \frac{\sigma(\degree_i)}{\langle \degree_i\rangle}=\sqrt{\frac{1}{\langle \degree_i\rangle} - \frac{\sum_{j\ne i}^Np^2_{i,j}}{L(\sum_{j\ne i}^Np_{i,j})^2}}.
\label{eq:coefVar}
\end{equation}
The quantity we are interested is the term 
\begin{equation}
\inversePartRat_i^{-1}=\sum_{j\ne i}^Np^2_{i,j}/\left(\sum_{j\ne i}^Np_{i,j}\right)^2
\label{eq:invPart}
\end{equation}
which is an \emph{inverse participation ratio}. This ratio measures the number of probability terms $p_{i,j}$ contributing effectively to Eq.~(\ref{eq:coefVar}). If only one probability term contributes, then $\inversePartRat_i=1$. If all the probability terms contribute in equal measure then $\inversePartRat_i =C$ for all $i$, where $C$ is a constant. The participation ratio gives a measure of the homogeneity of the statistical properties evaluated from the ensemble.

\subsection{Results}
We compare the models presented here with two of the most common null--models,  Newman--Girvan's model~\cite{newman2004finding} (NG) and restricted randomisation (RR). In Newman--Girvan's the probability that nodes $i$ and $j$ share a link is $\eProb_{i,j}=\degree_i\degree_j/(2L)$.  As the probability may exceed the value of $1$, the model is applicable if $k_ik_j>2L$ or $k_{max}<\sqrt{2L}$~\cite{boguna2004}. The restricted randomisation consist of reshuffling at random the links between the nodes, with the restriction that the degree of the nodes is conserved. If there is the extra restriction that two nodes cannot share more than one link, we refer to this model as the RR1 model. The RR1 model will produce correlated networks if the maximum degree is larger than the cut--off degree $k_{max}$.  If multiple links between two nodes are allowed but not self loops we called this model  RR2. In this case is possible to have multiple links between nodes so we expect that the nulls produced by this model will tend to be almost uncorrelated. They would be some correlation as it is expected that the number of self--loops for a node of degree $k$ is $k^2/(\langle k \rangle N)$~\cite{dorogovtsev2010lectures}, where $\langle k \rangle$ is the average degree, as in here self--loops are not allowed, then the null model could have some small bias in the correlations~\cite{boguna2004}.

\begin{figure}
\begin{center}
\subfigure[]{
\includegraphics[width=4.5cm]{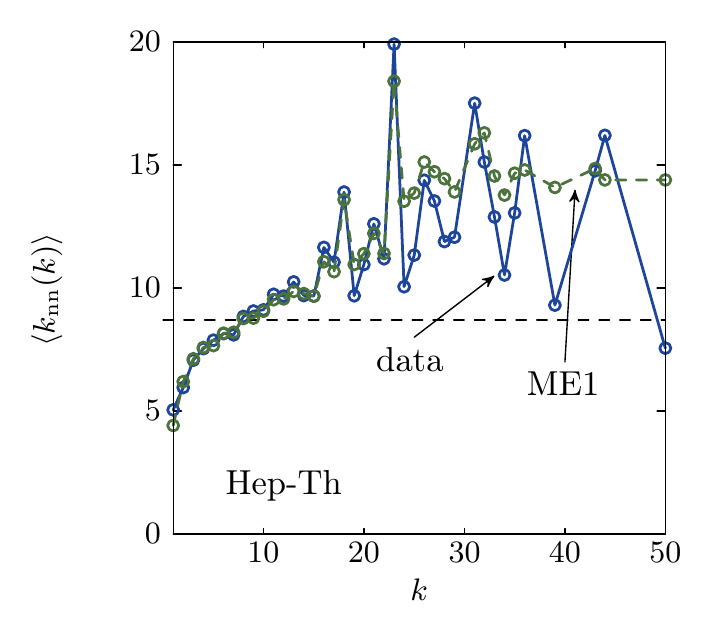}
}
\subfigure[]{
\includegraphics[width=4.5cm]{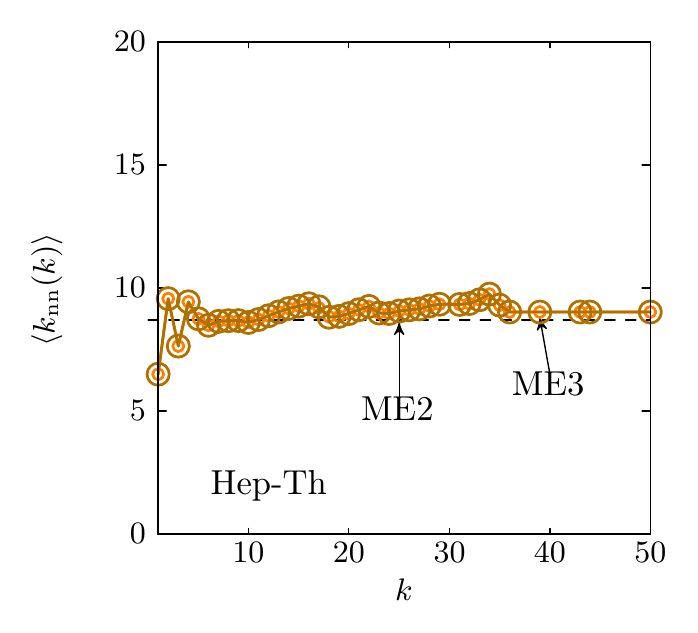}
}
\subfigure[]{
\includegraphics[width=4.5cm]{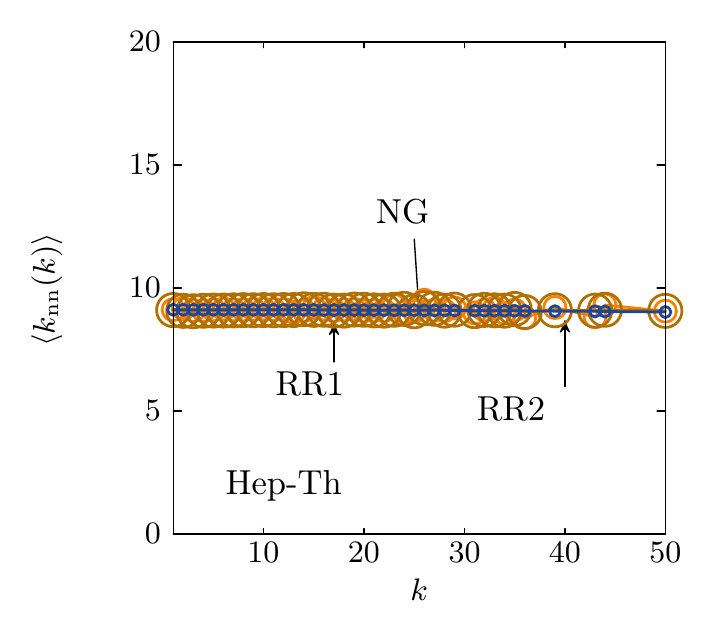}
}

\subfigure[]{
\includegraphics[width=4.5cm]{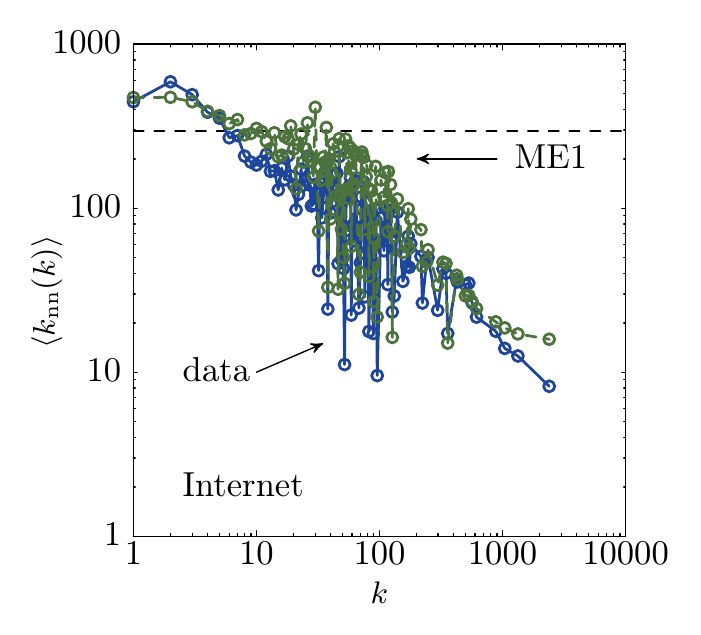}
}
\subfigure[]{
\includegraphics[width=4.5cm]{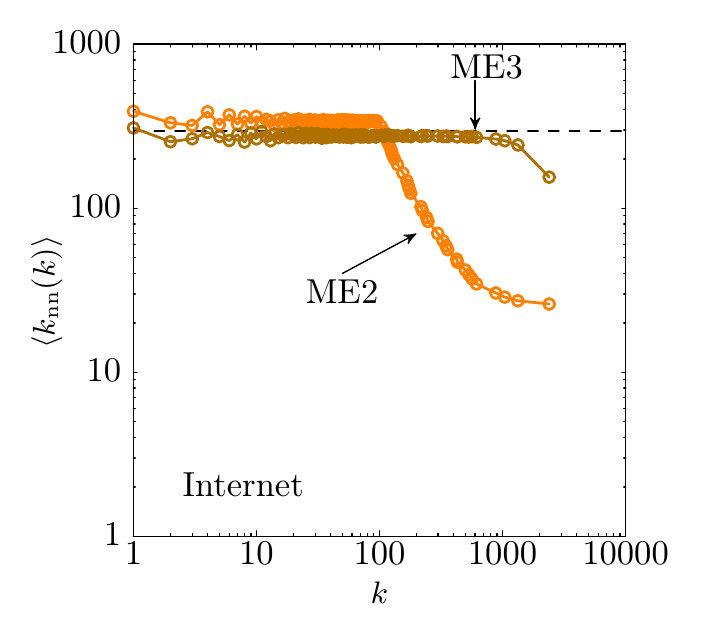}
}
\subfigure[]{
\includegraphics[width=4.5cm]{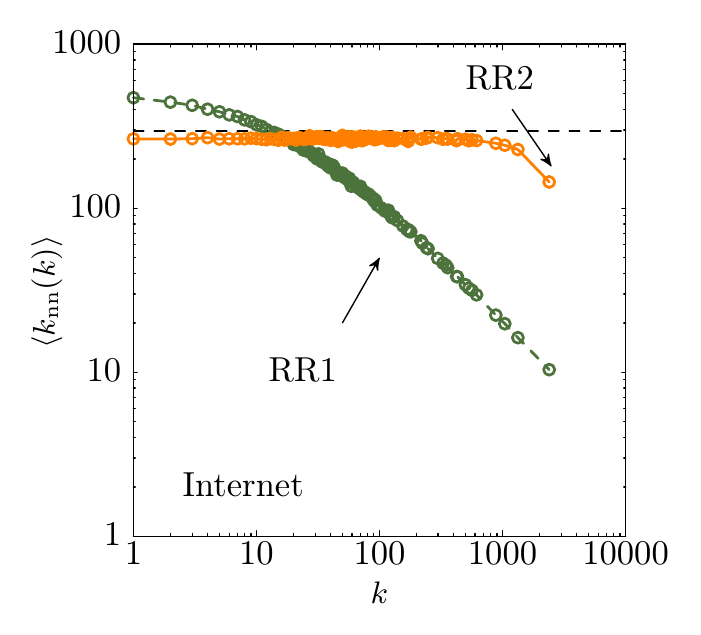}
}
\end{center}
\caption{\label{fig:degdegResults}  Comparison of the  average neighbours degree for the Hep--Th and AS Internet networks and their null--models. The original network and the null ME1 (left column). The entropy maximisation nulls ME2 and ME3 (middle column).  Newman--Girvan (NG), randomisation with (RR1) and without constraints (RR2) in the number of links between nodes (right column).  The dashed line shows the value of  $\langle k_{\rm nn} \rangle = {\langle k^2 \rangle}/{\langle k \rangle}$ which corresponds to a decorrelated network.
}
\end{figure}

Fig.~\ref{fig:degdegResults} shows the average neighbours degree  $\langle \averageDegDeg(k) \rangle$ obtained form the data and the different null models for the co--authorship of High Energy Physics (Hep--Th) and the AS--Internet networks. The Hep--Th network was chosen as it is assortative. The AS--Internet was chosen because the maximum degree is larger than the cut--off degree.

For the Hep--Th network the null model ME1  has a similar degree--degree correlation as the original network~(Fig.~\ref{fig:degdegResults}(a)). If we remove the constraint that there is, on average,  only one link between two nodes then,  as we do not have a degree cut--off,  both models ME2 and ME3 produce ensemble with similar degree--degree correlation~ (Fig.~\ref{fig:degdegResults}(b)). Which is also the case for the RR1, RR2 and NG models (Fig.~\ref{fig:degdegResults}(c)).

For the Internet, as in the previous example, the ME1 generates an ensemble with very similar degree--degree correlation as the original network (Fig.~\ref{fig:degdegResults}(d)). However, this network has a degree cut--off ($k_{c}\approx 260$) implying that if that, on average, only one link per pair of nodes is allowed then there is a degree--degree correlation between nodes. This is the case for ME2 and RR1 (Fig.~\ref{fig:degdegResults}(e)--(f)). There is a difference between the RR1 and the ME2 methods. In the ME2 multiple links are not allowed, the maximisation of the entropy produces an ensemble that minimises the degree--degree correlation but, due to the cut--off degree restriction, it fails to decorrelate  nodes of high degree. Compared this with the RR1,  which produces an ensemble where there are correlations at all degrees.

For the case that we allow more than one link between two nodes, the two models MM3 and RR2 produce almost uncorrelated networks. There still some correlation at high degrees, this correlation is due to restriction that  self--loops are not allowed. We do not present the results for the NG model as  $k_{max} >\sqrt{2L}$.

\begin{figure}
\begin{center}
\subfigure[]{
\includegraphics[width=5cm]{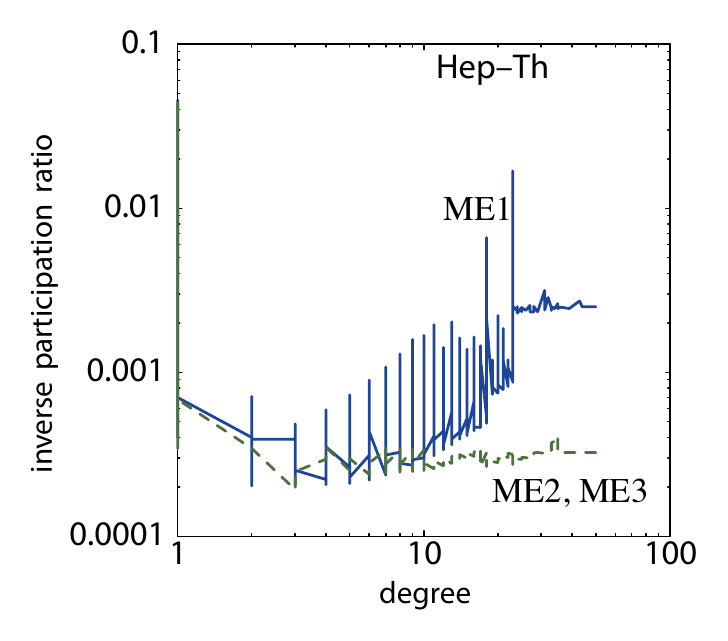}
}
\subfigure[]{
\includegraphics[width=5cm]{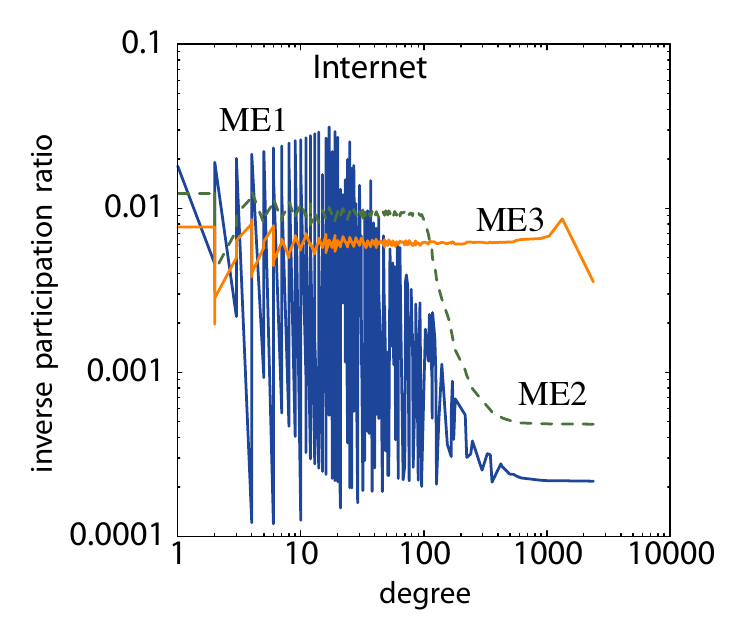}
}
\end{center}
\caption{\label{fig:invPartRat} Inverse participation ratio for the (a) High Energy Physics co--authors networks and the (b) AS Internet for the different MaxEnt ensembles.}
\end{figure}

Fig.~\ref{fig:invPartRat} shows the inverse participation ratio (Eq.~\ref{eq:invPart}) for the Hep--Th and AS Internet for the ME1, ME2 and ME3 ensembles.  Similarly as the results obtained using the average neighbours degree, the inverse participation ratio distinguishes if the ensemble produces correlated networks. In both networks the ME1 ensemble has a widely varying inverse participation ratio as by construction the ensemble is correlated. For the Hep--Th network the  probabilities $p_{i,j}$ obtained from ME2 and ME3 are such that terms $\sum_{i=1}^{N}p_{i,j}$ are almost homogenous, this is expected as by construction the MaxEnt method produces probabilities $p_{i,j}$ that are maximally neutral. This is not the case for the AS Internet, the existence of a cut--off degree and the constraints of no multiple links (ME2) or no self--loops (ME3) shows that the MaxEnt solutions, `push' towards homogenous contributions in the evaluation of $\sum_{i=1}^{N}p_{i,j}$ but this is not achieved due to the structural constraints. Nevertheless, this departure from homogeneity can be use to define the cut--off degree, in this case when the inverse participation ratio shows that the probability terms stop contributing in equal measure. For the case of the AS Internet, this happens $k_{cut}=102$ for  ME2 and $k_{cut}=1042$ for ME3.

\section{Communities and the ensembles}
A common use of null models is in the discovery of network communities~\cite{newman2004finding}. The communities are obtained by maximising the modularity function $Q$. If $a_{ij}$ is an entry of the adjacency matrix and $e_{ij}$ is the expected number of links between nodes $i$ and $j$ then the modularity is
\begin{equation}
Q = \sum_{i}^N\sum_{j}^N \left(a_{ij}-e_{ij} \right) \delta(g_i, g_j)
\end{equation}
where $g_i$ denotes the community that node $i$ belongs and $\delta(x,y)=1$ if $x=y$ and zero otherwise.
The most common null--model used in modularity maximisation is Newman--Girvan model, where $e_{ij}=\degree_i\degree_j/(2L)$. 

Squartini and Garlaschelli~\cite{squartini2011}  argued that there are several disadvantages using Newman--Girvan's null model in the discovery of communities via the modularity function, the main limitation is that the NG model is only feasible for networks where the maximum degree satisfies $k_{max}<\sqrt{2L}$. They proposed that a better approach is to use an $e_{ij}$ obtained using a maximum--likelihood (maximal entropy) method. Here we follow their suggestion and consider  $e_{ij} = Lp_{ij}$ where the probability $p_{ij}$ can be obtained from one of the null--models presented here. 

There are many techniques to obtain the maximisation of the modularity function $Q$~\cite{fortunato2010community}. Here we evaluate the community structure using the spectral method~\cite{newman2006}. This method partitions the network into two subnetworks. Then each subnetwork is partition again into two parts, and so on. This method creates a hierarchical partition of the network that can be represented with a dendrogram. The partition stops when the modularity is maximal or if it is not possible to split the network any further. Usually this method is stopped when the modularity is maximal but it is known that this does not imply that we obtain the best community structure as some networks have high modularity but poor community structure and other networks have low modularity and good community structure~\cite{massen2005identifying,karrer2007robustness}. We would apply the spectral method to our networks until is not possible to partition them any further, regardless if the modularity is maximal.

It is also known that the modularity methods are not deterministic as the communities obtained depend on the initial conditions used when starting the procedure~\cite{chakraborty2013,seifi2013,lancichinetti2011}. To take this dependancy into consideration we randomised the ranking of the nodes of equal degree. We evaluate the communities using 100 networks with randomised ranks  and then we evaluate how many times a pair of nodes appear in the same community.  The set of nodes that always appear in pairs and share a link are considered the core of the communities as they are independent of the ranking order and they must play a more relevant role in the structure of the community~\cite{chakraborty2013}.

\begin{figure}
\begin{center}
\subfigure[]{
\includegraphics[width=6cm]{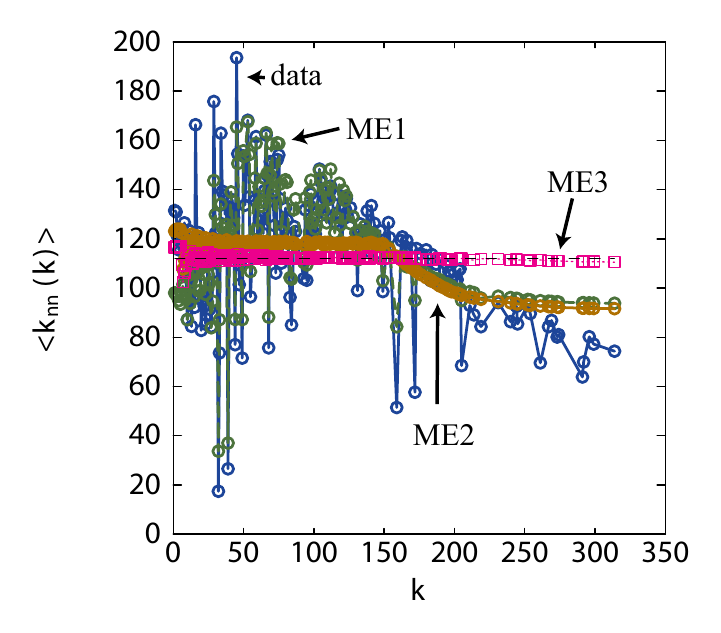}
}
\subfigure[]{
\includegraphics[width=6cm]{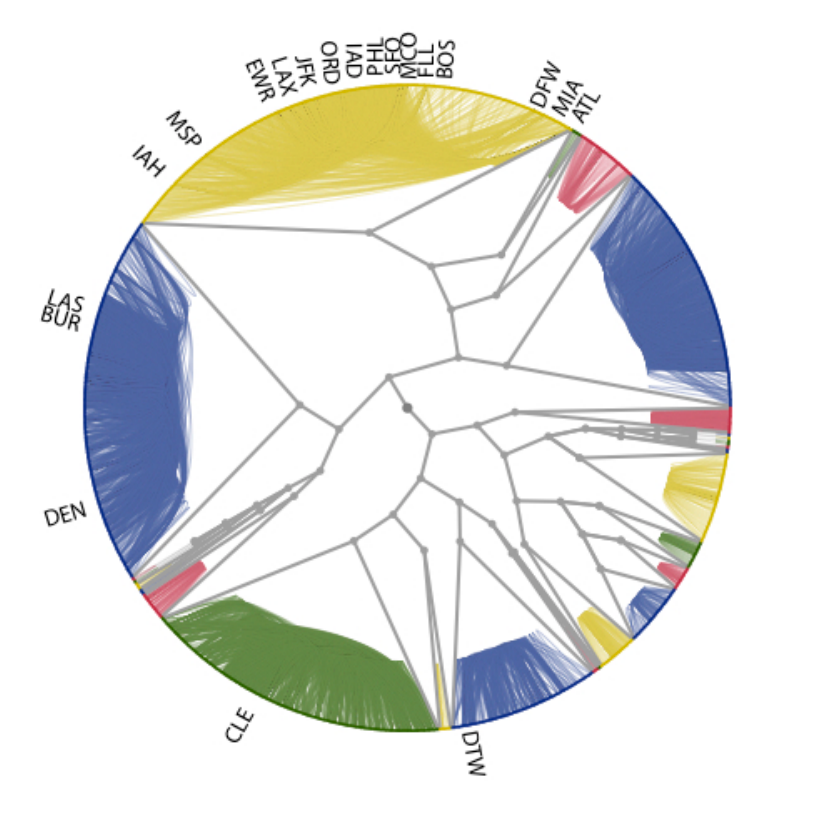}
}\\
\subfigure[]{
\includegraphics[width=6cm]{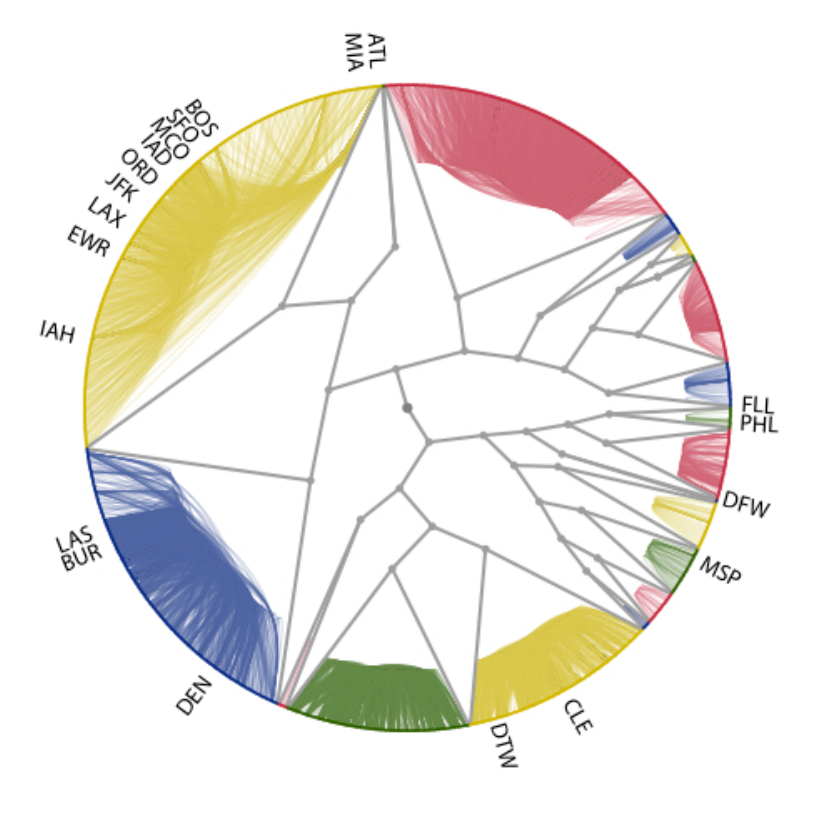}
}
\subfigure[]{
\includegraphics[width=6cm]{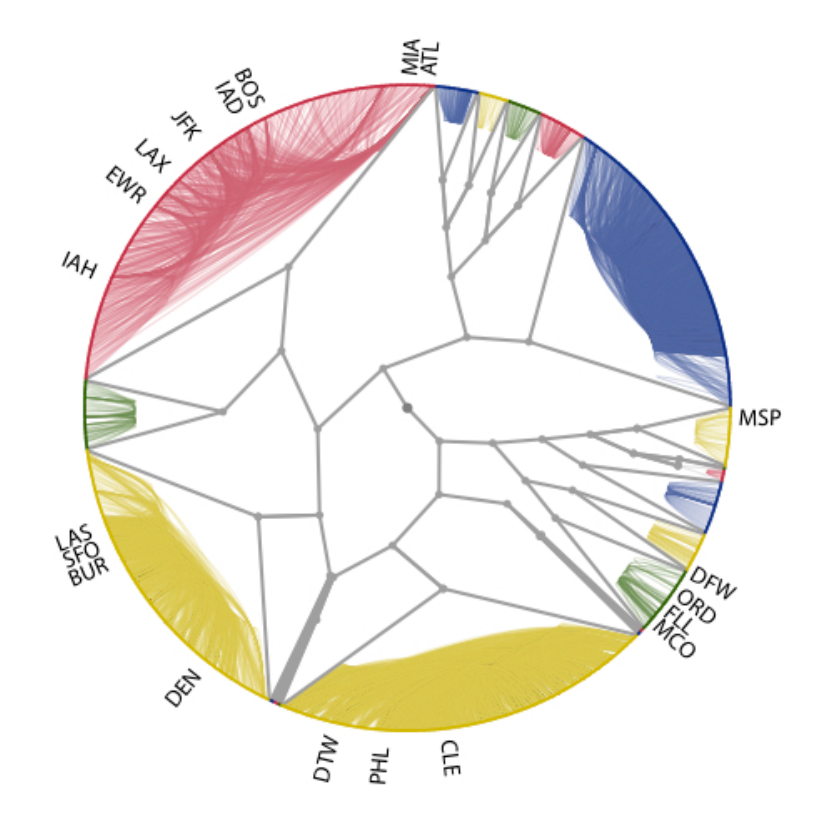}
}
\end{center}
\caption{\label{fig:airportCoinc} (a) The average nearest neighbours degree measured for the data (blue) and from the three null models (ME1:green, ME2:brown, ME3:pink).  Invariant communities obtained using different ensembles, (b) ME1, (c) ME2 and (d) ME3. The three letter are the acronyms of the airports, the full name can be found in http://www.transtats.bts.gov.
}
\end{figure}

As an example we evaluated the community structure of the US--airport dataset~\cite{opsahl2011anchorage} under the nulls ME1, ME2, and ME3. This network has a maximum degree of $k_{max}=312$ and a cut--off degree $k_c=180$ as a consequence, the ME2 will produce an ensemble with degree--degree correlations (Fig~\ref{fig:airportCoinc}(a)). ME1 generates correlated ensembles due to the conservation of the rich--club, and ME3 uncorrelated ensembles. 

Figure~\ref{fig:airportCoinc} (b)--(d) shows the invariant communities obtained from the different null models and their links represented with arcs joining them. The top 20 airports are labeled in the figure. These 20 airports have a degree greater than the cut--off degree, so the communities containing them would be different depending on the null used when defining them. The relationship between the communities given by the bipartition of the network is shown as a dendrogram in the figure. The large dot is the root node of the dendrogram.

The existence of correlations has an effect on the number of invariant communities and their composition. The ME1 case, where the ensemble has similar correlations to the real network, 15 of the top 20 airports belong to the same community. For the ME2 and ME3 ensemble,  the top 20 airports are more evenly distributed between the communities. 

In particular, LAS , BUR and DEN are always members of the same community independent of the ensemble used. Their membership to the same community is independent of the correlations. Other airports like CLE, DLW and PHL belong to the same community for the ME3 case. However if there are correlations due to the cut--off degree (case ME2) then CLE and DLW will be in the same community and, if we consider a correlated case (ME1), then all three airports belong to different communities.

From this example, it is clear that the existence of correlations in the null models has a major impact in defining the communities. What it is not clear is how to assess the effect that a change in the null--model's correlation has in the membership of the communities. It is known that a small change, for example the addition of one link, in the real network can have a drastic effect on the composition of the communities. For instance the resolution limit in community detection~\cite{fortunato2007resolution}, where the addition of one link between two well defined communities can be interpreted as the existence of a strong correlation between these communities and they are merged into one community. In this case, the null--model characteristics change very slightly with the new link, so the expected number of links between the communities is almost the same with and without this link, is the addition of a link in the real network that triggers the merger of the communities. To mitigate this behaviour we consider the case where the communities are defined using two ensembles. One ensemble captures the relevant properties of the real network, in this case the degree sequence and the degree--degree correlations. The real network is consider as a possible realisation from this ensemble. The other ensemble, which is the null model, has the same degree sequence as the real network but removes the degree--degree correlations, so there is no bias in the links connecting the nodes, this ensemble is used to evaluate, all things been equal,  the expected number of links between the communities

\subsection{Soft communities}
If $e^{(m)}_{ij}=Lp^{(m)}_{i,j}$ is the expected number of links between nodes $i$ and $j$ when considering the null model $m$, the communities are obtained from the modularity
\begin{equation}
Q = \sum_{ij}^N \left( \frac{e^{(1)}_{ij}-e^{(2)}_{ij}}{\sqrt{s^{(1)}_{ij} + s^{(2)}_{ij}}} \right) \delta(g_i, g_j)
\label{eq:myModularity}
\end{equation}
where $g_i$ denotes the community that node $i$ belongs, $\delta(x,y)=1$ if $x=y$ and zero otherwise. The term $s^{(m)}_{ij}=Lp^{(m)}_{i,j}(1-p^{(m)}_{i,j})$ is the variance of the mean for the null model $m$ and is introduced here as we are comparing the average given by two distributions. Notice that this normalisation is done between two pair of nodes not between the communities. As the constraints defining the ensembles are the averages of the degree and rich-club coefficient, we called the communities obtained from the maximisation of the modularity, soft--communities.
 
\begin{figure}
\begin{center}
\subfigure[]{
\includegraphics[width=6cm]{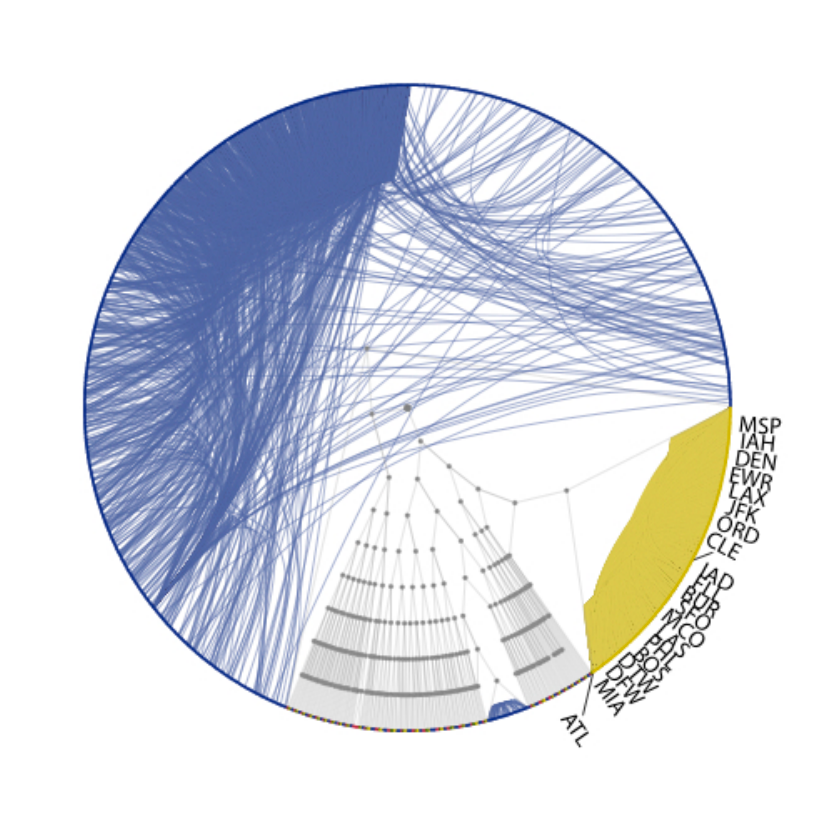}
}\hfill
\subfigure[]{
\includegraphics[width=9cm]{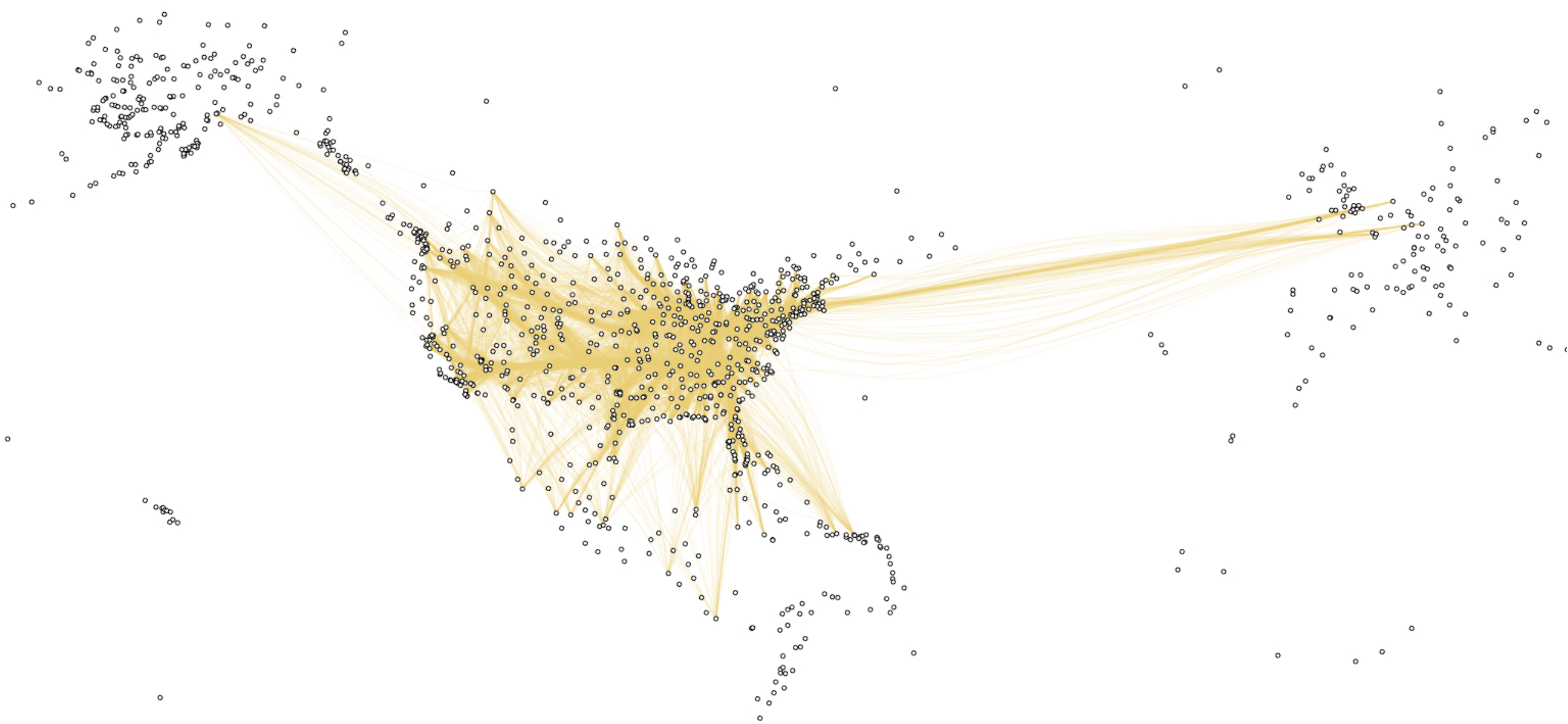}
}
\end{center}
\caption{\label{fig:airpWeighted} Communities defined by ensembles for the Airport network. (a) The network has three connected communities. One very densely connected community which contains the top 20 airports. (b) Geographical location of the densely connected community, this community also forms the core of the network.}
\end{figure}

Fig.~\ref{fig:airpWeighted}(a) shows the communities for the airports network when we compare the average number of links between the network with correlations given by ME1, as the correlated ensemble and ME2 and ME3 as the decorrelated ensembles. The use of ME2 or ME3 as the null gives almost identical communities so the figure shows the case of ME3. The communities were obtained using the spectral method.  The network has  three large communities, one of them contains all the top 20 airports. For the rest of the nodes each individual node is its own community. Using this `smooth' partition gives a structure similar to a  core--periphery. The core contains the top 20 airports which are densely connected. The periphery is all the other nodes which some form communities and some are isolated nodes. For this network the soft communities are not affected by the cut--off degree induced correlations. Fig~\ref{fig:airpWeighted}(b) shows the geographical location and links between the community that contains the 20 top airports.  

 As in the previous example, we evaluated the communities from 100 networks with randomised ranks  and measure how many times a pair of nodes appear in the same community. There was very little variation on the composition of the communities as a function of the ranking order, suggesting that these are stable communities.

\section{Conclusions}
Using the maximal entropy method produces null models that capture the correlation of the real networks and also can produce null models  with minimal bias. In the case of real networks with a  cut--off degree greater than the maximum degree, our method removes the bias for low degree nodes and minimises the bias for high degree nodes, contrasting with link-swapping randomisation methods which  introduce correlations at all degrees. For the maximal entropy method presented here it is easy to evaluate the cut--off degree where the removal of the bias is not possible, giving us a better understanding of the null model and hence of its constraints when the null model is used to define network structures.

Representing the network structure via ensembles allow us to define soft communities. The communities obtained from the ensembles are more robust to the non-deterministic property of the modularity maximisation algorithms and hence the determination of the network's communities. Defining the communities using these ensembles could take into consideration that the links describing the connectivity of the network could be incomplete or noisy.


\section{Appendix: Algorithm to evaluate the ensemble}
The entropy is
\begin{equation}
S = -\sum_{j=1}^N \sum_{i=1}^N p_{i,j} \log (p_{i,j}) = -2\sum_{j=1}^N \sum_{i=j+1}^{N} p_{i,j} \log (p_{i,j})
\end{equation}
as we are considering undirected networks with no self--loops. Using Eq.~(\ref{eq:mainResult}) 
\begin{eqnarray}
\label{eq:SimpEntro}
S&=&-2\sum_{i=1}^N\sum_{j=i+1}^{N}\left( \frac{w(i)(\degree_i-\linksRC_i)}{\sum_{n=1}^{j-1}w(n)(\degree_n-\linksRC_n)} \frac{\linksRC_j}{L} \right) \log \left(  \frac{\stepFunct(i)(\degree_i-\linksRC_i)}{\sum_{n=1}^{j-1}\stepFunct(n)(\degree_n-\linksRC_n)} \frac{\linksRC_j}{L}\right)
\nonumber\\
&=&-2\sum_{i=1}^N\sum_{j=i+1}^{N}\left( \frac{F(i)}{G(j)} \frac{\linksRC_j}{L} \right) \log \left(  \frac{F(i)}{G(j)} \frac{\linksRC_j}{L}\right)\\
\nonumber
&=&-2\left( \sum_{i=1}^N\frac{F(i)}{L}\log \left(\frac{F(i)}{L} \right) \left[ \sum_{j=i+1}^{N} \frac{\linksRC_j}{G(j)}\right]
+\sum_{i=1}^N\frac{F(i)}{L}\left[ \sum_{j=i+1}^{N} \frac{\linksRC_j}{G(j)}\log \left(  \frac{\linksRC_j}{G(j)}\right)\right] \right)\\
&=& -2\left(\sum_{i=1}^N\frac{F(i)}{L}\log \left(\frac{F(i)}{L} \right)A(i)
+\sum_{i=1}^N\frac{F(i)}{L}B(i) \right)
\end{eqnarray}

where $F(i) = \stepFunct(i)(\degree_i-\linksRC_i)$, $G(i) = \sum_{j=1}^{i-1}F(j)$, $A(i)= \sum_{j=i+1}^{N} \frac{\linksRC_j}{G(j)}$ and $B(i)=\sum_{j=i+1}^{N} \frac{\linksRC_j}{G(j)}\log \left(  \frac{\linksRC_j}{G(j)}\right)$. 

\subsection*{Algorithm}
For the null models where the sequence $\{\linksRC_1,\dots, \linksRC_N\}$ is not given, this sequence is obtained by exploring the space of networks, searching for the ensemble with maximal entropy. The search is done by choosing two nodes at random, say node $i$ and $j$, if the first node satisfies $\linksRC_i<\degree_i-1$ and the second node satisfies $\linksRC_j >0$ then we increase $\linksRC_i$ by one link and decrease $\linksRC_j$ by one link. This simple procedure is to satisfy the constraint $\sum_{i=1}^N\linksRC_i=L/2$. Every time that a $\linksRC_i$ is changed the probabilities $\{\eProb \}$ are evaluated using Eqs.~(\ref{eq:mainResult}) and (\ref{eq:stepFunct}). The entropy of the network is evaluated from Eq.~(\ref{eq:SimpEntro}).

We used Simulated Annealing and a greedy algorithm to find $\{\linksRC_r\}$, we noticed that both algorithms gave very similar results, so we present only the greedy algorithm.

\begin{algorithmic}[1]
\REQUIRE $\{\degree_i\}$ sorted in decreasing order
\REQUIRE $\{\linksRC_i\}$. If this sequence is not given generate it at random with the condition $\linksRC_i \le \degree_i$
\STATE $S_{old} \leftarrow S$, evaluate the entropy $S$
\STATE $\delta S \leftarrow 1$
\WHILE{$\delta S \ne 0$} 
\STATE Select nodes $i$ and $j$ at random, where $i\ne j$
\IF{ $\linksRC_i<\degree_i$ and $\linksRC_i<i-1$ and $\linksRC_j\ge 1$} 
\STATE $\linksRC_i \leftarrow \linksRC_i +1$
\STATE $\linksRC_j \leftarrow \linksRC_j -1$
\STATE $S_{new} \leftarrow S$, evaluate the new entropy
\STATE $\delta S \leftarrow S_{new}-S_{old}$, evaluate the change in the entropy
\STATE $S_{old} \leftarrow S_{new}$
\IF{$\delta S > 0 $} 
\STATE $\linksRC_i \leftarrow \linksRC_i -1$
\STATE $\linksRC_j \leftarrow \linksRC_j +1$
\ENDIF
\ENDIF
\ENDWHILE

\COMMENT {At the end of the while loop we obtain the sequence $\{\linksRC_r \}$ which gives the network with minimal entropy}
\end{algorithmic}
Lines 6-7 change the connectivity of the rich--club coefficient via $k\linksRC_r$, lines 11-14 reject the new connectivity if the entropy has not decreased and reset the values of the rich--club coefficient. 

For the case that the  constraints are $\linksRC_i\le \degree_i$ only line 5 from the above algorithm needs to be modified to
\begin{algorithmic}
\IF {$\linksRC_i<\degree_i$ and $\linksRC_j\ge 1$} 
\STATE $\ldots$
\ENDIF
\end{algorithmic}

To evaluate the entropy 
\begin{algorithmic}[1]
\REQUIRE $\degree_m$ and $\linksRC_m$ for $m=1,\ldots, N$
\STATE $\stepFunct(1)=1$
\STATE $G(1) \leftarrow w(1)(\degree_1-\linksRC_1)$
\FOR{$m=2$ to $N$}
\STATE $\stepFunct(m) \leftarrow \stepFunct(m-1)*G(m-1)/(G(m-1)-\linksRC_m*\stepFunct(m))$
\STATE $G(m) \leftarrow G(m-1)+\stepFunct(m)*(\degree_m-\linksRC_m)$
\ENDFOR
\STATE $B(1)=0$
\FOR{$m=N$ to $2$}
\STATE $A(m) \leftarrow \linksRC_m/G(m)$
\STATE $B(m) \leftarrow B(m-1)+(\degree_m/G(m)*\log(\degree_m/G(m)))$
\ENDFOR
\FOR{$m=1$ to $N$}
\STATE $a \leftarrow 0$
\IF{$\stepFunct(m)*(\degree_m-\linksRC_m) \ne 0$}
\STATE $a \leftarrow \stepFunct(m)*(\degree_m-\linksRC_m)/L *\log(\stepFunct(m)*(\degree_m-\linksRC_m)/L)$
\ENDIF
\STATE $S(m) \leftarrow S(m-1) + a*A(m) + \stepFunct(m)*(\degree_m-\linksRC_m)*B(m)/L$
\ENDFOR
\end{algorithmic}
\newpage
\bibliographystyle{unsrt}

%

\end{document}